\documentclass[acmsmall,screen,nonacm]{acmart}

\AtBeginDocument{%
  }

\usepackage{subcaption}
\usepackage{algorithm}
\usepackage{algorithmic}
\usepackage{listings}

% ...
\definecolor{primary500}{HTML}{7c4f86}
% ...
\definecolor{secondary500}{HTML}{565d9b}
% ...
\definecolor{accent500}{HTML}{3f713a}
% ...
\definecolor{aux500}{HTML}{904961}
% ...
\definecolor{aux2500}{HTML}{855811}

\lstdefinestyle{tinyCPP}{
    language=C++,
    basicstyle={\tiny\ttfamily},
    keywordstyle=\color{primary500},,
    stringstyle=\color{accent500},
    commentstyle=\color{secondary500},
}

\lstdefinestyle{tinyMLIR}{
    sensitive=false,
    basicstyle={\tiny\ttfamily},
    stringstyle=\color{primary500},
    string=[b]",
    commentstyle=\color{secondary500},
    comment=[l]{//},
    keywordstyle=[1]\color{secondary500},
    keywordstyle=[2]\color{primary500},
    keywordstyle=[3]\color{accent500},
    keywordstyle=[4]\color{aux500},
    keywordstyle=[5]\color{aux2500},
    alsoletter={\%},
    escapechar=\$,
    keywords=[1]{sdfg, arith, memref, func, scf, return, math},
    keywords=[2]{
        \%a, \%b, \%c, \%true, \%result, \%r, \%i, \%e, \%d, \%2, \%1, \%0, \%10, \%c0, \%c0f
    },
    keywords=[3]{f32, f16, i32},
    keywords=[4]{@double, @main},
    keywords=[5]{true},
}

\lstdefinestyle{tinyWASM}{
    sensitive=false,
    basicstyle={\tiny\ttfamily},
    stringstyle=\color{primary500},
    string=[b]",
    commentstyle=\color{secondary500},
    comment=[l]{//},
    keywordstyle=[1]\color{secondary500},
    keywordstyle=[2]\color{primary500},
    keywordstyle=[3]\color{accent500},
    keywordstyle=[4]\color{aux500},
    alsoletter={\$},
    keywords=[1]{block, func},
    keywords=[2]{
        \$a, \$b, \$c, \$double
    },
    keywords=[3]{i32, f32},
    keywords=[4]{global, local},
}

\newcommand{\macsection}[1]{\textbf{\textit{#1.}}~~}

\begin{document}

\title{MLIR-Forge: A Modular Framework for Language Smiths}

\author{Berke Ates}
\email{berke.ates@inf.ethz.ch}
\orcid{0000-0003-0242-3640}
\affiliation{%
  \institution{ETH Zurich}
  \city{Zurich}
  \country{Switzerland}
}

\author{Philipp Schaad}
\email{philipp.schaad@inf.ethz.ch}
\orcid{0000-0002-8429-7803}
\affiliation{%
  \institution{ETH Zurich}
  \city{Zurich}
  \country{Switzerland}
}

\author{Timo Schneider}
\email{timo.schneider@inf.ethz.ch}
\orcid{0000-0002-4884-3934}
\affiliation{%
  \institution{ETH Zurich}
  \city{Zurich}
  \country{Switzerland}
}

\author{Alexandru Calotoiu}
\email{acalotoiu@inf.ethz.ch}
\orcid{0000-0001-9095-9108}
\affiliation{%
  \institution{ETH Zurich}
  \city{Zurich}
  \country{Switzerland}
}

\author{Torsten Hoefler}
\email{htor@inf.ethz.ch}
\orcid{0000-0002-1333-9797}
\affiliation{%
  \institution{ETH Zurich}
  \city{Zurich}
  \country{Switzerland}
}


\begin{abstract}
    Optimizing compilers are essential for the efficient and correct execution of software across various scientific fields.
    Domain-specific languages (DSL) typically use higher level intermediate representations (IR) in their compiler pipelines for domain-specific optimizations.
    As these IRs add to complexity, it is crucial to test them thoroughly.
    Random program generators have proven to be an effective tool to test compilers through differential and fuzz testing.
    However, developing specialized program generators for compiler IRs is not straightforward and demands considerable resources.
    We introduce MLIR-Forge, a novel random program generator framework that leverages the flexibility of MLIR, aiming to simplify the creation of specialized program generators.
    MLIR-Forge achieves this by splitting the generation process into fundamental building blocks that are language specific, and reusable program creation logic that constructs random programs from these building blocks. This hides complexity and furthermore, even the language specific components can be defined using a set of common tools.
    We demonstrate MLIR-Forge's capabilities by generating MLIR with built-in dialects, WebAssembly, and a data-centric program representation, DaCe --- requiring less than a week of development time in total for each of them. Using the generated programs we conduct differential testing and find $9$ MLIR, $15$ WebAssembly, and $774$ DaCe groups of bugs with the corresponding program generators, after running them until the rate of new bugs stagnates.
\end{abstract}

\begin{teaserfigure}
  \center
  \includegraphics[width=0.8\textwidth]{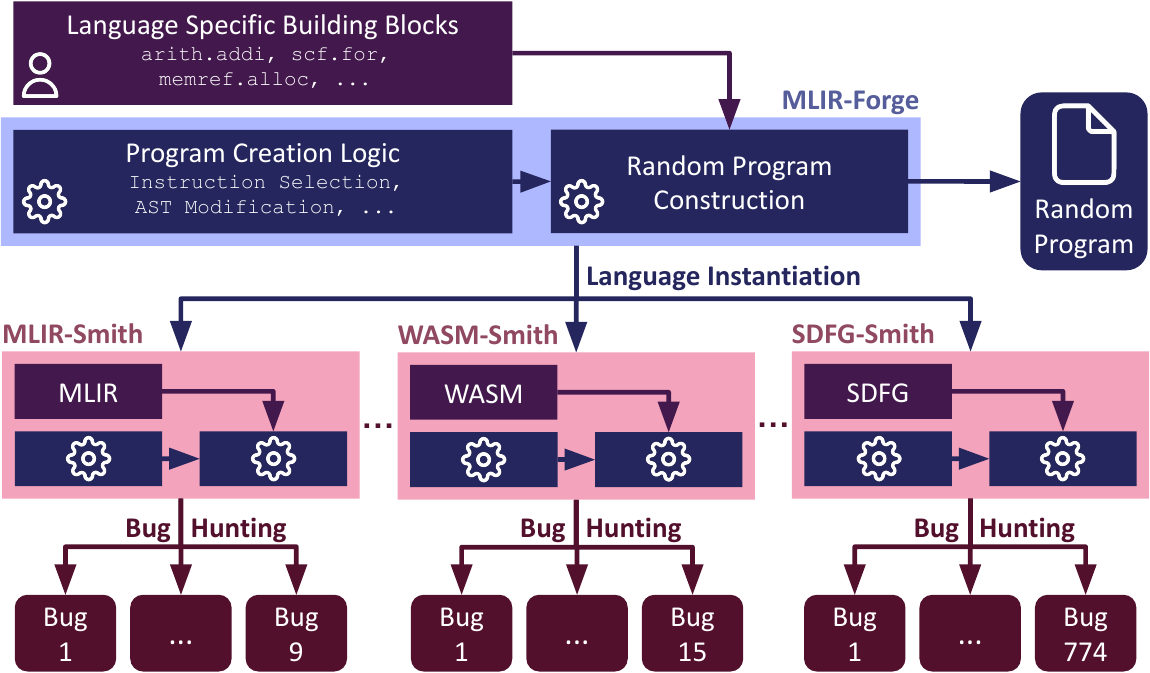}
  \caption{Overview of MLIR-Forge.}
  \label{fig:posterchild}
\end{teaserfigure}


\maketitle

\newpage
\section{Introduction}

To achieve good performance in the current post-Moore era, both hardware and software are increasingly more specialized and optimized for specific tasks~\cite{DBLP:journals/corr/abs-1807-04188, 10.1145/2830772.2830791}.
%
Following this trend of specialization, programming languages have also evolved to cater to specific domains with so called Domain-Specific Languages (DSLs)~\cite{Muller:HASE05,DBLP:journals/corr/abs-1302-6333}.
DSLs are tailored for specific tasks or industries and make it easier for experts in a field to work with computational tools. 
The vast majority of DSLs and general-purpose languages utilize intermediate representations (IRs) in their compiler pipelines to perform general and purpose-specific optimizations.
%
General-purpose programming languages often use several higher-level IRs before moving to a more general low-level IR. This approach allows language-specific optimizations to easily be implemented in a modular fashion, such as in the Soot framework, where three IRs are used to effectively optimize Java bytecode~\cite{10.1007/3-540-46423-9_2}.

MLIR~\cite{mlir} (Multi-Level Intermediate Representation) is a key development in this area. It simplifies the process of creating a new IR by offering a framework capable of handling many IRs and promoting reuse through a provided pass infrastructure, tools for diagnostics, testing facilities, and much more.
%
Thanks to MLIR, creating and experimenting with new IRs has become easier and quicker. This has encouraged more research in the fields of machine learning, quantum computation, and code optimization through the use of specialized IRs~\cite{10.1145/3469030, chelini2022mom, DBLP:journals/corr/abs-2008-08272, hu2023tpumlir, mccaskey2021mlir}.

Early testing of such custom IRs is vital to prevent costly bug fixes later on and ensure trust in experimental results~\cite{rajamanickam2019software}.
%
Random program generators are valuable tools for finding compiler bugs. Tools such as Csmith~\cite{10.1145/1993316.1993532} and DeepSmith~\cite{10.1145/3213846.3213848} demonstrate how many compilers could crash or produce incorrect code despite receiving valid input programs.
%
%
Handwritten tests usually follow structured patterns with sensible inputs due to subconscious biases introduced by the human developers writing them.
In contrast, random program generators can create unusual and unexpected programs containing patterns far outside of what is typically seen in sensible code. While both are valid, the unpredictable nature of randomly generated code can challenge a compiler in unique ways. This makes program generators effective at revealing complex bugs that standard tests might miss.
%
However, developing a program generator for a specific IR can be time-consuming and expensive. This can take away from the savings of using MLIR for IR development.

We introduce MLIR-Forge, a program generator framework in MLIR that
%
separates this task into program generation and language specification, and automates the program generation procedure. This limits the developers' responsibility to specifying the fundamental language building blocks -- e.g., defining the IR operations specific to the language. We summarize our approach in Fig.~\ref{fig:posterchild}.
MLIR-Forge provides a set of common tools to minimize the effort required to implement a program generator, making testing more affordable and common -- a goal aligned with MLIR's philosophy.
Our approach facilitates code reuse, allowing users to leverage the work of other MLIR-Forge-based program generators.

We demonstrate the use of MLIR-Forge by creating a program generator for a subset of MLIR's built-in dialects, and one for a custom dialect for program dataflow optimizations, the SDFG dialect~\cite{mlir-dace}. We show that even IRs not originally designed for MLIR can benefit from MLIR-Forge by combining the MLIR program generator with translators to produce and test WebAssembly~\cite{10.1145/3140587.3062363} code.

In summary, our main contributions are:
\begin{itemize}
    \item Separating random program generator development into program construction and language specification by defining an abstraction layer between them.
    \item Automating the program construction process and reducing developer responsibility to language specification alone.
    \item Facilitating reuse between program generators through sharing language specifications for shared language features.
    \item Demonstration of the framework with a proof-of-concept implementation in MLIR, used to implement three program generators, each in less than one week, and uncovering $798$ groups of bugs.
\end{itemize}

\section{IR Abstraction}\label{sec:2_irabstraction}
To build a random program generator framework capable of producing a variety of IRs, a common IR abstraction is essential.
This abstraction provides a unified way to describe individual IRs within the framework.
The following section outlines the abstraction, which offers a standardized expression for instructions across different IRs.

\subsection{Operation Model}
\begin{figure}[h]
    \centering
    \includegraphics[width=.99\linewidth]{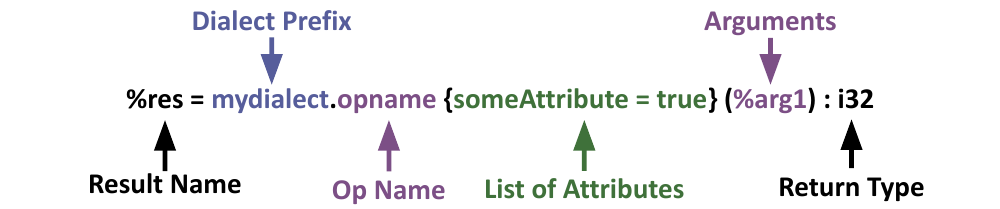}
    \caption{Overview of an operation.}
    \label{fig:2_mlir_op}
\end{figure}

Since instructions in IRs can have arbitrary semantics, a fixed instruction set is insufficient to represent a wide variety of IRs.
We address this challenge by using an abstracted view on operations, as depicted in Fig.~\ref{fig:2_mlir_op}.

These operations can take any number of operands or arguments and produce zero or more results. Operations are grouped into dialects, which are namespaces for operations and types. To identify the dialect of operations, they are prefixed with the dialect's name, which allows different dialects to share the same name for operations with differing application-specific semantics, such as arithmetic and vector addition.

Operations may also contain attributes which represent static information that can change the behavior of the operation. For example, a rounding operation might have an attribute that sets the rounding method. Operations can also use arguments which are direct results from other operations. While the type of an operation usually tells us about its results, it can also give information about its arguments.

\begin{figure}[h]
    \centering
    \begin{subfigure}[b]{0.45\linewidth}
        \centering
        \includegraphics[width=\linewidth]{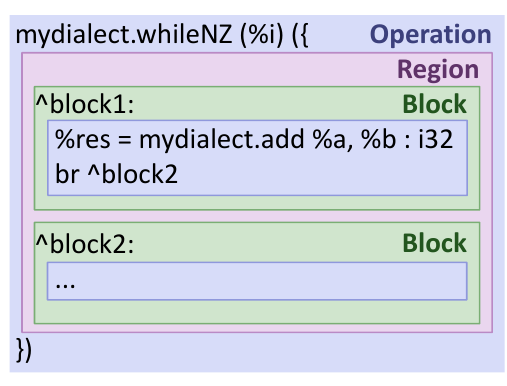}
        \caption{Illustrations of the nested operation structure.}
        \label{fig:2_mlir_regions}
    \end{subfigure} 
    \hfill
    \begin{subfigure}[b]{0.47\linewidth}
        \centering
        \includegraphics[width=\linewidth]{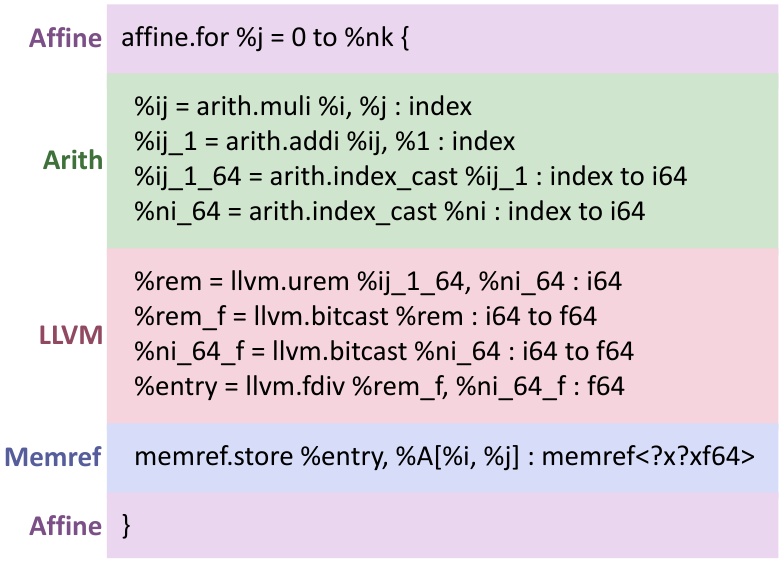}
        \caption{Dialects of different abstraction levels coexisting.}
        \label{fig:2_mlir_dialects}
    \end{subfigure}
    \caption{Illustrations of MLIR concepts.}
    \label{fig:2_mlir_concepts}
\end{figure}

Operations follow a hierarchical or nested design which allows them to contain multiple regions. Each region is made up of basic blocks, containing a list of operations. This design enables us to represent even (structured) control flow instructions as operations, as shown in Fig.~\ref{fig:2_mlir_regions}.
%
By grouping operations into namespaced dialects, we can utilize multiple dialects in the same program, as seen in Fig.~\ref{fig:2_mlir_dialects}. This design allows IR developers to build on existing dialects, focusing on what is new in their IR. Specialized dialects can therefore target specific areas in hardware and software more effectively.

\subsection{MLIR}
\begin{figure}[h]
    \centering
    \includegraphics[width=.99\linewidth]{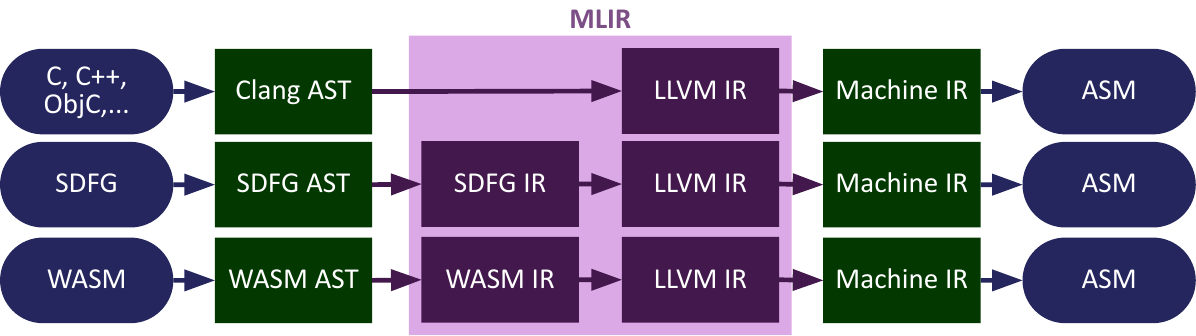}
    \caption{Illustration of the IR specialization trend and the unification through the MLIR framework.}
    \label{fig:2_mlir_overview}
\end{figure}

This abstract operation model is practically implemented in the MLIR framework, making a prime target for creating the program generator framework. One of its standout features is its ability to handle IRs at any level of abstraction. As shown in Fig.~\ref{fig:2_mlir_overview}, MLIR can be used to chain a variety of such IRs to make use of different abstraction levels and their strengths at different stages of the compilation process.
%
%
MLIR's design is both flexible and modular, making it straightforward to create different IRs. It also comes with tools to analyze and debug these custom IRs, avoiding reimplemenation. Using MLIR's infrastructure, users can implement translators that translate any IR into one that is MLIR-defined and vice versa. 
This brings up an interesting point: Could we combine existing random program generators with a translator to create random programs for a specific IR?

\begin{figure}[h]
    \centering
    \includegraphics[width=.99\linewidth]{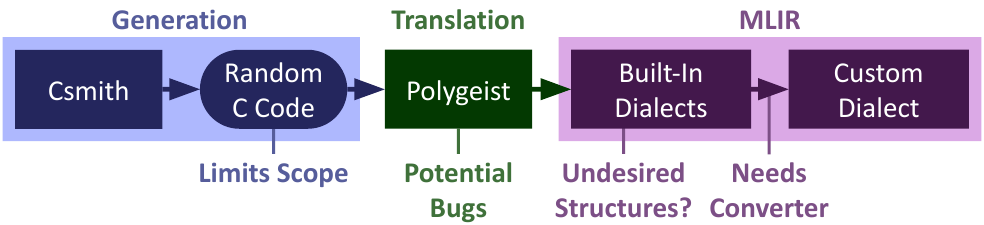}
    \caption{Illustration of an example Csmith and Polygeist pipeline with potential drawbacks.}
    \label{fig:2_csmith_polygeist}
\end{figure}

One approach to creating program generators for MLIR-defined IRs is to use existing generators with a translator. For example, we could use Csmith~\cite{10.1145/1993498.1993532} to generate randomized C code and then translate it into MLIR using Polygeist~\cite{polygeistPACT}, as illustrated in Fig.~\ref{fig:2_csmith_polygeist}.
%
%
Relying on such a method has its drawbacks. One of the challenges is the reliability and accuracy of the translator. Adding a translator introduces another layer where bugs might arise, making testing more complex. This is not merely speculative. In our tests with Csmith and Polygeist, we observed that Polygeist crashes on programs generated by Csmith. Additionally, as the goal is to assess the IR, translator-related problems are counterproductive distractions.

We are also limited by the capabilities of the source language. For instance, C does not support nested functions, i.e. declaring a function inside of a function, so any IR that does would be left untested. Further, we cannot control how the translator handles the input code, which could lead to unwanted code structures. Lastly, translators emit MLIR built-in dialects. To get custom dialects, we would need to additionally implement a converter from built-in dialects to custom dialects.
%
Even if one designs a custom translator to directly produce the desired custom IR dialect, that still requires significant work and potentially introduces more bugs.
%
Given these challenges, it is more effective to generate random programs directly within MLIR. This approach can utilize the whole MLIR framework. Specifically, rather than building a specific IR generator, we can develop a framework, which provides IR designers with tools to implement random program generators for their specific IR with minimal overhead.

\section{Building a Program Generator Framework}\label{sec:3_approach}
\begin{figure}[h]
    \centering
    \includegraphics[width=0.99\linewidth]{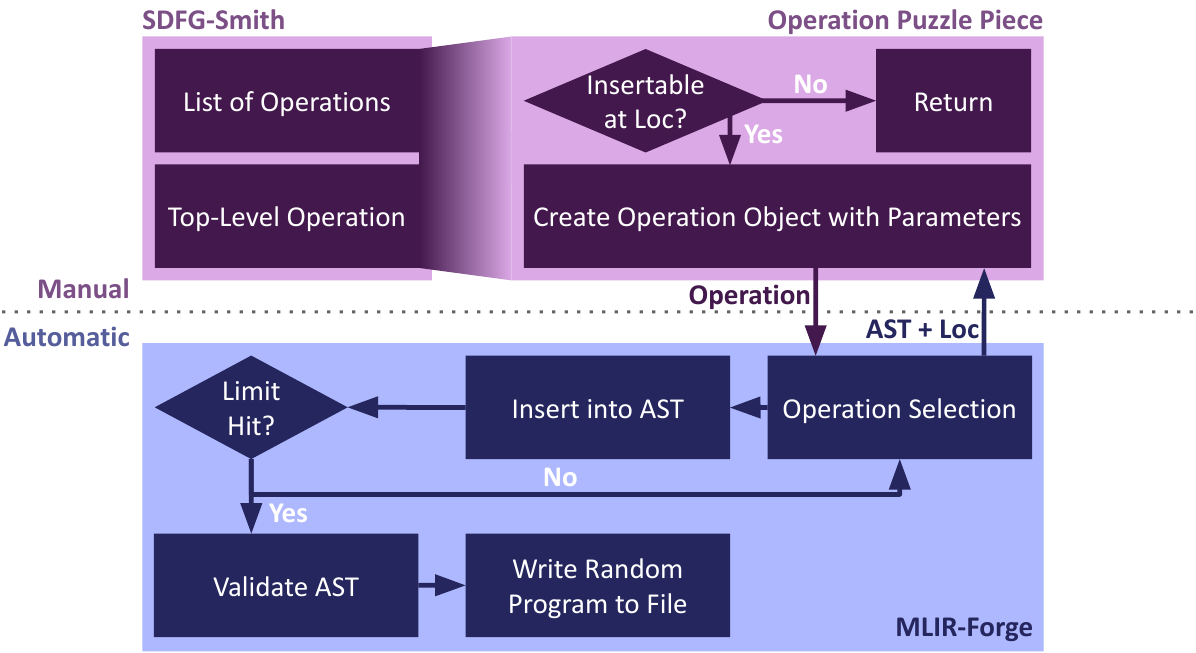}
    \caption{Illustration of the components of a random program generator and the abstraction through MLIR-Forge. We consider the example of a random program generator for the SDFG dialect, but the components would be the same regardless of the IR.}
    \label{fig:3_flow}
\end{figure}

Our approach to a program generator framework consists of breaking down the language specification into fundamental building blocks, akin to jigsaw puzzle pieces.
Here we detail our framework abstraction, which is illustrated in Fig.~\ref{fig:3_flow}.

\subsection{Defining Puzzle Pieces}
The introduction of these puzzle pieces changes the generation process from the perspective of a user or IR developer to a process where the generator \emph{inserts} certain program elements at the right time, and the developer only tells the generator how program elements can be inserted.

Each puzzle piece is tied to a specific feature of the programming language, such as an operation. When a puzzle piece is invoked, it receives the existing structure of the current program along with the location of the next insertion. With this context, the puzzle piece must then determine if it can insert the language feature into the program at the given location. If the conditions do not allow for insertion, the puzzle piece simply returns. Otherwise it proceeds to generate an object of the language feature, filled with all necessary parameters, such as arguments and types, and returns it ready for insertion. 
Since many parameters share common constraints among the different puzzle pieces, we also provide tools to assist in their specification.

To allow for this, the language specification can be thought of as a series of individual generators, or Gens for short.
Each such generator defines a specific piece of the puzzle in the language specification.
For instance, each operation the language supports has an associated operation generator, or OpGen for short, which describes the insertion process into an existing program.
Going back to the analogy of a jigsaw puzzle, this describes the shape of the piece, which represents an operation.
Similarly, types are represented by type generators, or TypeGens, that serve the same purpose.

As an example we consider the SDFG IR~\cite{mlir-dace}, which has different operations, including arithmetic addition. Using our approach, we would implement an OpGen (i.e. puzzle piece) specifically for this arithmetic addition. When this OpGen is used, it would check if there are two integers available at the current location. If two integers are available, the OpGen would create an addition object and use these integers as the arguments for that addition.

This shift to an insertion-centric perspective allows for the reuse of Gens across different program generators. For instance, arithmetic expressions, common across many languages, can be described once and then reused in any other representation where they share the same specifications.
This both promotes efficiency in constructing program generators, and reduces the potential for bugs.
%
In other words, the generator is now able to piece together a series of puzzles at random from the shape of the puzzle pieces (i.e., the language specification), and a developer's wishes (i.e., configurations and flags) alone.
The framework knows the rules of playing a jigsaw puzzle independent of what pile of pieces the user hands it.

\subsection{Abstracting and Partitioning the Process}
The second part in our partitioned approach involves an automatic framework that uses the puzzle pieces to construct a random program.
%
The framework starts by picking a puzzle piece at random and asks it to provide the associated language feature. If the puzzle piece can provide this feature, the framework adds it to the Abstract Syntax Tree (AST). After each insertion, the framework checks if it has reached the limit for creating the program. If it has not reached the limit, it continues the process with another puzzle piece. Once the limit is reached, the framework makes sure the AST follows the rules of the language specification and saves the random program to a file.

Using the SDFG IR as an example again, the framework might choose an addition OpGen at first. If there are no integers available for the addition operation, the OpGen would signal a failure. The framework might continue to select a constant OpGen. After inserting the constant operation, there are now integers available, which would allow the framework to successfully insert an addition operation.

%
By generalizing these components into reusable modules, a program generator framework can alleviate most of the development burden for a program generator by extracting the recipes for how these components function from the language specification.
%
%
%
With this abstraction, the primary responsibility left to the user is to provide the puzzle pieces and a top level piece, hiding details such as the selection of puzzle pieces and modification of the AST.

\section{MLIR-Forge}
MLIR-Forge serves as a practical example of such a general random program generator framework, where certain aspects of MLIR are leveraged to hide the complexity of the program construction process from the IR developer.
%

In MLIR-Forge, a random program generator constructs programs one operation at a time, from top to bottom.
It starts by picking an operation at random from a selection pool of available operations and then calls its related OpGen. This OpGen tries to create and place the operation. If it is placed successfully, MLIR-Forge moves the insertion point, which is the location where new operations are inserted, one step down and repeats the process. 
If an operation cannot be created, for instance because values with the required types are unavailable, the OpGen informs MLIR-Forge of the failure. MLIR-Forge then removes the unsuccessful OpGen from its selection pool and picks another operation. This selection process continues until either the pool is exhausted or a successful OpGen is found. Once this happens, the selection pool is reset, including re-adding previously unsuccessful OpGens.
The selection process halts if either no successful OpGen can be found or a program length limit is reached.
Control flow instructions do not require special treatment since MLIR represents them as singular operations.





The foundation of MLIR-Forge consists of three main components: An interface for operations and types, a Command-Line Interface (CLI) framework, and an IR builder. These components are outlined and further elaborated in the following.

\subsection{Interfaces and CLI Framework}
We integrated interfaces for operations and types, enabling the addition of OpGens and TypeGens to their corresponding operations and types.
Recognizing that many dialects utilize LLVM's TableGen with MLIR's Operation Definition Specification (ODS) backend, we additionally implemented operation and type traits, which add our interfaces to the generated operation and type declarations.

These interfaces do not offer a default implementation for OpGens or TypeGens, acting as a check against missed Gen implementations. Successful OpGens return a pointer to the newly added operation, allowing for convenient tracking of the changes made. If an OpGen fails, it simply returns a null pointer. On the other hand, TypeGens return a type object, which is then utilized by the OpGens, rather than directly inserting into the IR.

We additionally introduce a generalized CLI framework, which is the main access point for MLIR-Forge. The user can build a customized CLI by passing their preferred dialects and a top-level operation to the CLI framework. It interprets user flags and modifies the configuration based on them. These flags allow users to set the seed, load and dump the configuration, and specify the output file. Once the flags are processed, the CLI framework initiates the program generation by creating a module operation, which is an implicit top-level operation in MLIR, and populating its block using the IR builder's utility function.

\subsection{IR Builder}
The purpose of the IR builder is to provide OpGens a convenient way to interact with the IR.
The IR builder draws from MLIR's OpBuilder, which already offers tools for IR navigation and the insertion of new operations, blocks, and regions. The IR builder overrides MLIR OpBuilder functions that move the insertion point of new operations in order to prevent the insertion point from being moved upwards. It also enhances the operation creation function to first check constraints before creating and inserting the desired operation. This function returns an operation pointer if successful, and a null pointer otherwise. This allows OpGens to confirm the success of their operation creation and communicate the outcome back to the IR builder.
%
The IR builder comes with utility functions that allow for sampling of values and types and gathering all available values and types in the current insertion point of the IR. This is done by moving upwards through the IR from the insertion point, collecting values and types along the way. The traversal stops when it hits the root operation or an operation that is isolated from the outer scope. 
The collector utilities return the gathered values and types, while the sampling functions uniformly pick a value from the collected choices.


The IR builder also provides tools for randomly creating operations and filling blocks with operations. This setup supports a recursive approach, allowing OpGens to produce operations as part of their own creation. Additionally, the interface design lets OpGens call the OpGen of another operation directly when a particular operation is needed.
%
The IR builder incorporates a nested config class that establishes various settings, such as the seed, maximum depth of nested regions, and operation sampling probabilities. This class is adaptable, enabling users to add their own configurations and apply them in their OpGens.
%
When initializing, the IR builder gathers all OpGens by verifying them against the interface. Operations with a zero probability are omitted for efficiency. 

\subsection{Challenges}\label{sec:4_challenges}
Generating entire program regions brings about a set of challenges. The result types of a region depend on the available values inside it. Thus, these result types can only be determined after the region has been fully generated, with the terminator operation, which marks the end of a code block, defining them in retrospect. For example, the func dialect provides the \texttt{func.func} operation to define a function. \texttt{func.func} can return values that are defined in its body with \texttt{func.return}. The types of these return values have to match the result types of the body region. Therefore, the result types of the body region are set by \texttt{func.return}'s OpGen instead of \texttt{func.func}'s OpGen.

While MLIR-Forge allows for arbitrary terminators to be sampled and created, by default they are excluded from the operation selection pool. We assume that parent operations will insert the necessary terminators. This is based on the understanding that regions typically know their required terminators, which are often few in number. For example, \texttt{func.func} expects \texttt{func.return}, \texttt{scf.while} expects \texttt{scf.condition} and \texttt{scf.yield}, basic blocks expect \texttt{cf.br}, \texttt{cf.cond\_br} or \texttt{cf.switch}.

Certain operations might have specific return type requirements for their regions. For example, the \texttt{scf.while} operation needs a boolean value in its condition region. Ensuring these types are present is a challenge, since MLIR-Forge is unaware of the types that can be created. While one could consider querying all potential types for decision-making, dialects can define infinite sets of types. For instance, the \texttt{memref} type can have any number of dimensions. Dialects might additionally have composite types made up of other types. Given these factors, MLIR-Forge cannot continuously create operations in regions hoping for the desired types to appear, as the generation process must terminate. Thus, MLIR-Forge does not assume any knowledge of the types it can create.

This limitation is relatively minor for two main reasons.
Firstly, despite being technically possible, it is uncommon for an operation to have a region, which always returns a specific type. Regions are generally intended for flexibility and restricting the return type would diminish their adaptability. Secondly, in our study of a subset of MLIR's built-in dialects, we identified only two operations, \texttt{scf.while} and \texttt{omp.atomic.update}, that have such constraints.


\subsection{Operation Frequencies}
One of MLIR-Forge's aims is to ensure that the frequencies of operations in the generated programs align with the frequencies specified in the configuration. Given this, we were interested in understanding how these constraints impact the frequency of operations. Specifically, we determined the likelihood of generating an \texttt{scf.while} operation, considering the constraints mentioned in \S\ref{sec:4_challenges}.

Let us consider there are $N$ operations before the \texttt{scf.while} operation and the condition regions of \texttt{scf.while} contains $K$ operations. Both $N$ and $K$ adhere to a geometric distribution with a probability of $p_g$. The likelihood of an operation producing the necessary boolean is represented as $p_{bool}$. To simplify the calculation, we assume all operations have an equal chance of being chosen and operations that are present do not influence the probability of other operations.

The probability distribution for the sum $S = N + K$ is:

\begin{displaymath}
\begin{split}
P(S = s) &= \sum^{s}_{n = 0} P(N = n) \times P(K = s-n) \\
&= \sum^{s}_{n = 0} p_g \times (1-p_g)^n \times p_g \times (1-p_g)^{s-n} \\
&= \sum^{s}_{n = 0} p_g^2 \times (1-p_g)^s = (s+1) \times p_g^2 \times (1-p_g)^s
\end{split}
\end{displaymath}

The probability of having at least one boolean value for a fixed $N$ and $K$ is $P(\#bool \geq 1) = 1 - (1-p_{bool})^{N+K}$. 
Using these, the probability of generating an \texttt{scf.while} operation given that the \texttt{scf.while} operation was chosen is

\begin{displaymath}
\begin{split}
&P(\textrm{\texttt{scf.while} generated}\ |\ \textrm{\texttt{scf.while} chosen}) \\
&=\sum^{\infty}_{s=0} (1 - (1-p_{bool})^{s}) \times P(S = s) \\
&=\sum^{\infty}_{s=0} (1 - (1-p_{bool})^{s}) \times (s+1) \times p_g^2 \times (1-p_g)^s
\end{split}
\end{displaymath}

For our prototype with $90$ operations, where only one operation (\texttt{arith.constant}) can unconditionally produce a boolean value, and using $p_{bool} = 1/90$ and $p_g = 0.2$, the computed probability is approximately $P(\textrm{\texttt{scf.while} generated}\ |\ \textrm{\texttt{scf.while} chosen}) \approx 0.0833 = 8.33\%$.
%
To validate this, we generated $10,000$ random programs using MLIR-Smith, and counted the occurrences of \texttt{scf.while}. The observed frequency was $8.05\%$, which is consistent with the theoretical prediction.

\section{Forging a Smith}
To construct a smith using MLIR-Forge, users need to use our MLIR fork equipped with MLIR-Forge.
A custom dialect mirroring the specific programming language or IR in question is required. 
The dialect should encompass operations found in the language or IR. 
The most efficient method to declare these operations is through the use of MLIR's ODS, which automates the generation of the required boilerplate C++ code.
%
Operations within this dialect must implement the \texttt{GeneratableOp} interface, providing them with a \texttt{generate()} function. This function, accepting an IR builder, either yields a pointer to an operation or, if insertion at the current location is not feasible, returns a null pointer. If insertion is possible, the function should gather any necessary parameters, construct an operation object, and then return it. A practical example is shown in Fig.~\ref{fig:4.5_addi_opgen}, which shows the OpGen of \texttt{arith.addi}.

\begin{figure}[h]
    \centering
    \begin{minipage}[t]{0.99\linewidth}
        \centering
\begin{lstlisting}[style=tinyCPP]
Operation *arith::AddIOp::generate(GeneratorOpBuilder &builder) {
  Type resultType = builder.getI32Type();
  llvm::Optional<Value> lhs = builder.sampleValueOfType(resultType);
  llvm::Optional<Value> rhs = builder.sampleValueOfType(resultType);

  if (!lhs.has_value() || !rhs.has_value())
    return nullptr;

  OperationState state(builder.getUnknownLoc(), "arith.addi");
  arith::AddIOp::build(builder, state, lhs.value(), rhs.value());
  Operation *op = builder.create(state);

  return op;
}
\end{lstlisting}
        \caption{Simplified OpGen of the integer addition in the arithmetic dialect.}
        \label{fig:4.5_addi_opgen}
    \end{minipage}
\end{figure}

%
Once all \texttt{generate()} functions are implemented, the next step involves establishing an entry point for the smith, often accomplished by populating a registry with all dialects involved and invoking the CLI framework. Typically, the framework defaults to treating the module operation as the top-level, but this can be customized.
%
At this stage, a smith capable of generating the dialect and offering CLI configuration via flags is realized. If the target IR is the MLIR dialect, no further action is required. Otherwise, a straightforward translator is needed to convert the MLIR-defined syntax to that of the target IR. Given MLIR's adaptability, it can be reconfigured to closely resemble the syntax of the target IR.

While our approach offers flexibility, it also has its limitations. 
It is the user's responsibility to implement safety measures, such as preventing division by zero.
Because user-defined operations can have arbitrary semantic meanings, safety cannot be automatically provided. 
For the same reason, ensuring the determinism of operations is up to the user. For example, users must initialize memory before accessing it to avoid non-deterministic behavior.

\section{Evaluation}
\begin{figure}[h]
    \centering
    \begin{subfigure}[b]{0.3\linewidth}
        \centering
\begin{lstlisting}[style=tinyMLIR]
func @double(%a: memref<10xi32>) 
{

  %2 = arith.constant 2
  %0 = arith.constant 0
  %1 = arith.constant 1
  %10 = arith.constant 10

  scf.for %i = %0 to %10 step %1
  {
  
    %e = memref.load %a[%i]
    
    %d = arith.muli %e, %2
    
    memref.store %d, %a[%i]
    
  }
  
  return
  
}
\end{lstlisting}
        \caption{MLIR}
        \label{fig:5_ir_overview_mlir}
    \end{subfigure} 
    \hfill
    \begin{subfigure}[b]{0.47\linewidth}
        \centering
        \includegraphics[width=\linewidth]{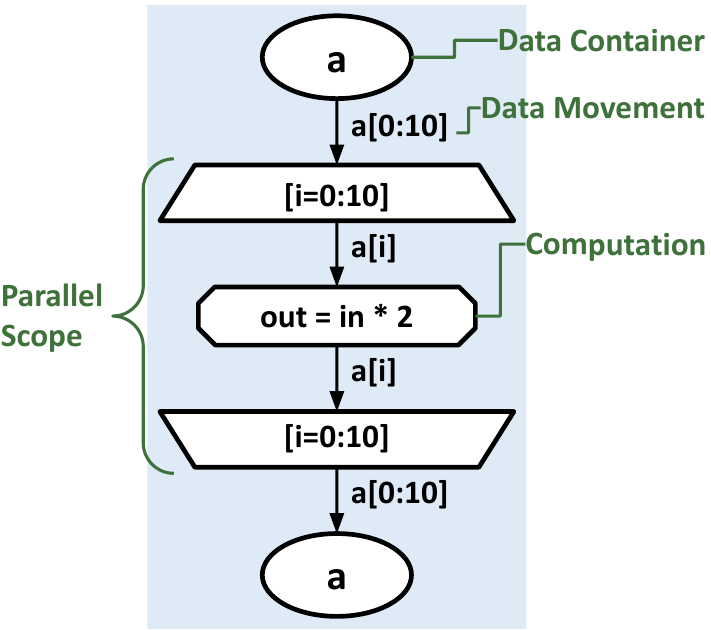}
        \caption{SDFG}
        \label{fig:5_ir_overview_sdfg}
    \end{subfigure}
    \hfill
    \begin{subfigure}[b]{0.2\linewidth}
        \centering
\begin{lstlisting}[style=tinyWASM]
(func (param i32) 
  (local i32 i32)
  loop
    local.get 0
    local.get 1
    i32.const 2
    i32.shl
    i32.add
    local.tee 2
    local.get 2
    i32.load
    i32.const 1
    i32.shl
    i32.store
    local.get 1
    i32.const 1
    i32.add
    local.tee 1
    i32.const 10
    i32.ne
    br_if 0
end)
\end{lstlisting}
        \caption{WASM}
        \label{fig:5_ir_overview_wasm}
    \end{subfigure}
    \caption{The three IRs used in our evaluation express the same doubling function in vastly different ways.}
    \label{fig:5_ir_overview}
\end{figure}

Using MLIR-Forge we developed three random program generators, targeting conceptually very different IRs: MLIR, SDFG (Stateful DataFlow multiGraph) IR~\cite{dace}, and  WebAssembly~\cite{10.1145/3062341.3062363} (WASM).
%
MLIR is an imperative IR, designed for representing a wide range of abstraction levels within a single framework. This makes it suitable for both high-level, domain-specific operations and low-level, hardware-specific instructions.
SDFG IR, on the other hand, is a data-centric IR. It represents programs as graphs where nodes are computational tasks and edges signify data dependencies. This approach is particularly effective for optimizing data movement and memory usage, crucial in high-performance and parallel computing environments.
WASM is a stack-based IR, optimized for efficient execution and compact representation. Its stack-based nature, where instructions operate on a stack of data values, makes it highly efficient for web applications, ensuring portability and compactness across different platforms.
Fig.~\ref{fig:5_ir_overview} illustrates the fundamental differences between these IRs by providing the same doubling function expressed in each IR.

Generating random programs for SDFGs presents its own unique set of challenges, primarily centered around accurately modeling and managing data dependencies and execution order. In SDFGs, the data flow is not just a sequence of operations but a complex network of dependencies, where the state and transformation of data at each node can have cascading effects throughout the graph. Randomly generating such graphs requires an understanding of data dependencies and the ability to create a coherent flow of data that respects the order of execution of the program. This involves ensuring that each node correctly processes and passes data to subsequent nodes, and that the overall graph structure remains valid and executable.

In the process of randomly generating WASM programs, a significant challenge is ensuring valid stack operations. As a stack-based language, WASM requires that each operation aligns with the stack's current state, either consuming or adding values to it. Random generation must, therefore, maintain a coherent stack state throughout the program. This involves accurately predicting the stack's status at every point, a task that becomes complex with control flow changes and function calls. 

By using conceptually different IRs with unique constraints, we demonstrate the ease with which robust program generators can be implemented for both established and custom program representations.
We demonstrate how this helps program testing efforts by using generated programs to perform differential fuzz testing, uncovering bugs in the IR's optimizer and compiler passes.
%

\subsection{SDFG-Smith}
DaCe~\cite{dace} is a data-centric optimization and code-generation framework, which uses the SDFG (Stateful DataFlow multiGraph) IR to represent programs from the data movement perspective. This is accomplished by expressing programs as directed acyclic graphs which orchestrate data as it flows through a series of computations. The DaCe ecosystem offers the SDFG dialect~\cite{mlir-dace}, which is capable of representing the SDFG IR in MLIR.
The first random program generator produces the SDFG dialect, which we process through a translator to obtain the SDFG IR.
%
%
%
We perform differential fuzz testing on the optimizing framework and compiler infrastructure around the SDFG IR using the randomly generated programs. SDFG-Smith produces programs which take one or more input argument and then passes them to a fuzzer to perform differential testing. We pass the programs to two fuzzers: AFL++~\cite{AFLplusplus-Woot20}, which aims to improve testing accuracy by maximizing branch coverage, and FuzzyFlow~\cite{schaad2023fuzzyflow}, a differential input fuzz-tester for the SDFG IR.
We stop our testing when the rate of novel bugs seems to stagnate.

\begin{figure}[h]
    \centering
    \includegraphics[width=\linewidth]{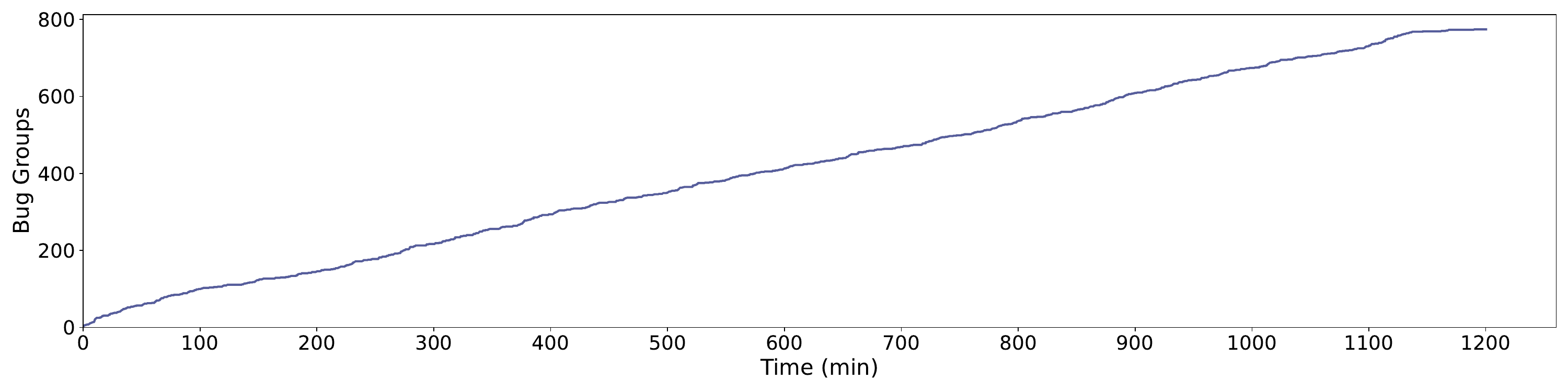}
    \caption{Comparison of bugs groups, using our grouping method, to testing time.}
    \label{fig:5_time_vs_bug}
\end{figure}

Since multiple inputs can trigger the same bug, we group them by taking the error message and computing the MD5 hash.
Because some error messages contain line numbers of the input program, we conservatively replace all numbers in the error message by a fixed character.
Fig.~\ref{fig:5_time_vs_bug} depicts the total number of groups we identified using our method across the runtime of our differential tester.
We generated $14,681$ programs over a period of $20$ hours and identified $774$ bug groups. In $1,154$ programs, DaCe's optimization passes reached our imposed time limit of $10$ seconds.
While a comprehensive analysis of each bug falls outside the scope of this work, we have collaboratively identified three significant bugs with the DaCe development team, which will be fixed in a future release.

\begin{figure}[h]
    \centering
    \includegraphics[width=.99\linewidth]{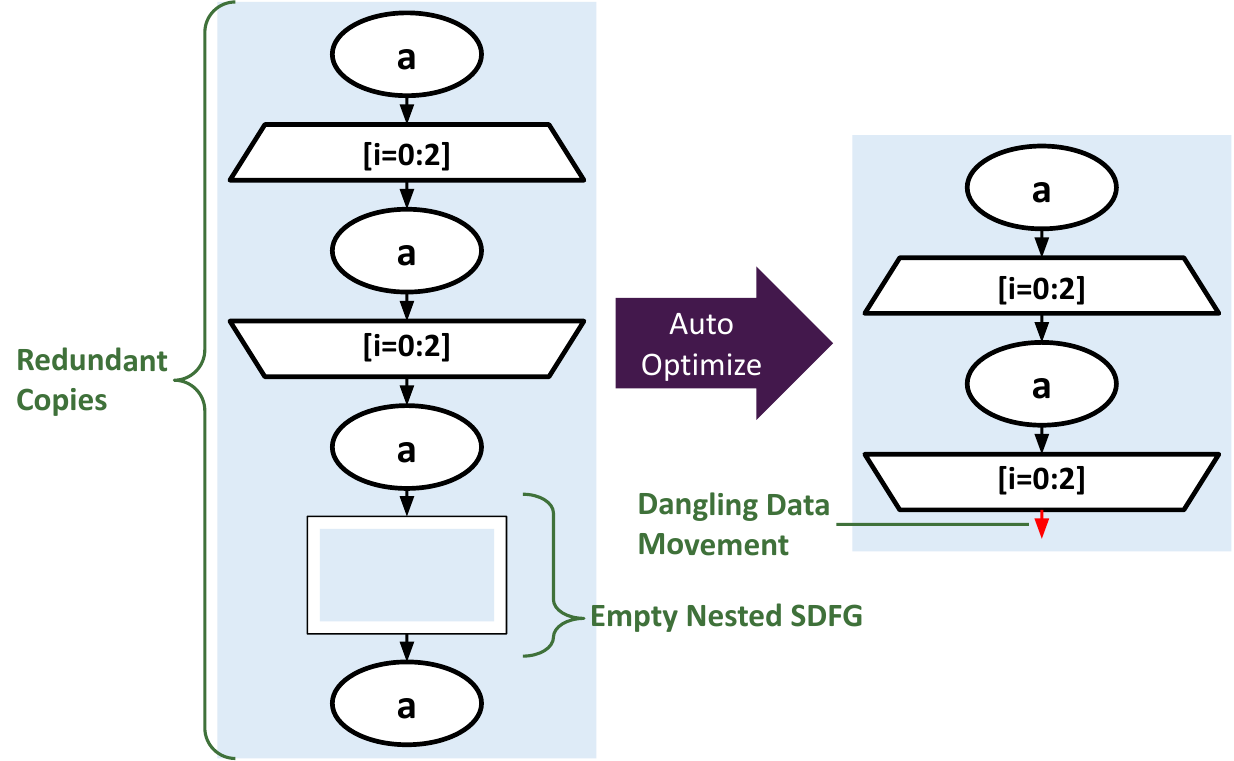}
    \caption{Simplified SDFG generated by SDFG-Smith revealing DaCe's auto optimize bug.}
    \label{fig:5_dace_bug}
\end{figure}

Fig.~\ref{fig:5_dace_bug} illustrates one of these bugs. The simplified input SDFG only performs redundant copies, which can be eliminated entirely from a data-centric point of view. The input SDFG also contains an empty nested SDFG, which is a subroutine that does not perform any operation, essentially functioning as a \texttt{NOP}. After optimizing using DaCe's auto-optimizer, we obtain a faulty SDFG. The auto-optimizer successfully recognizes the redundant copies at the bottom of the SDFG, but fails to identify the parallel scope as dead dataflow. 
This leads to the erroneous removal of the lower nodes, leaving the parallel scope with a dangling data movement with no target.

\subsection{MLIR-Smith}
MLIR-Smith, our second random program generator, generates MLIR code using a subset of built-in dialects, specifically SCF, Arith, Math, Memref, and Func.
%
The process begins with the generation of these dialects, followed by optimization (\texttt{-cse --canonicalize --symbol-dce --loop-invariant-code-motion --inline}) of the resulting MLIR code.
The next step involves lowering both the unoptimized and optimized MLIR to the LLVM dialect, and then to LLVM IR~\cite{10.5555/977395.977673} using MLIR's \texttt{mlir-opt} and \texttt{mlir-translate} tools.
We compile the LLVM IR using LLC (\texttt{-O0 --relocation-model=pic}) and Clang (\texttt{-O0 -fPIC -march=native}), checking for errors or crashes during all stages. 
%
The compiled binaries are run with a set time limit to identify any runtime issues.
If both binaries either reach the time limit or exit with the same code, the test is considered successful. 
If only one binary exceeds the time limit, it suggests a potential discrepancy in termination behavior between the two binaries. Since a binary might simply run significantly slower, further analysis is required to confirm a change in termination. Differing exit codes between the binaries clearly signal a behavioral change, indicating a bug.

%
Using this process we identified a case where the unoptimized code did not terminate, whereas the optimized version did. Fig.~\ref{fig:5_mlir_loop} shows a simplified version of the unoptimized code. The difference in termination is caused by an infinite while loop that was removed during optimization. Discussions with MLIR developers revealed that the specifications do not currently cover this case. They suggested two options: define such behavior as undefined or consider the loop side-effect-free, allowing for its elimination during dead code elimination.
We discovered similar cases involving deallocating the same memory multiple times and deallocating stack-allocated memory.
We generated $23,065$ programs over a period of $10$ hours and identified $9$ bug groups.

\begin{figure}[h]
    \centering
    \begin{minipage}[t]{0.47\linewidth}
        \centering
\begin{lstlisting}[style=tinyMLIR]
func @main() {
  %true = arith.constant true

  scf.while () : () -> () {
    scf.condition(%true) 
  } do {
    scf.yield
  }

  return
}
\end{lstlisting}
        \caption{Simplified MLIR code generated by MLIR-Smith revealing the absence of specifications for non-terminating while loops.}
        \label{fig:5_mlir_loop}
    \end{minipage} 
    \hfill
    \begin{minipage}[t]{0.47\linewidth}
        \centering
\begin{lstlisting}[style=tinyMLIR]
func @main() -> i32 {

  %c0f = arith.constant 0.0 : f16
  %c0 = arith.constant 0 : i32

  %1 = math.fpowi %c0f, %c0 : f16, i32
  %2 = math.absf  %2 : f16
    
  return %c0 : i32
  
}
\end{lstlisting}
        \caption{Simplified program generated by WASM-Smith crashing \texttt{emcc} with \texttt{-O3} written in MLIR.}
        \label{fig:5_wasm}
    \end{minipage}
\end{figure}

\subsection{WASM-Smith}
Building on the capabilities of MLIR-Smith, our third random program generator extends the generation process to WebAssembly~\cite{10.1145/3062341.3062363} (WASM) code.
The output generated by MLIR-Smith is lowered to LLVM IR using MLIR's \texttt{mlir-opt} and \texttt{mlir-translate} tools. We use the emscripten~\cite{10.1145/2048147.2048224} compiler frontend \texttt{emcc} to obtain WASM code. Using the WebAssembly Binary Toolkit WABT we can translate the WASM code into human-readable WebAssembly text format (WAT).
This setup highlights MLIR-Forge's ability to produce complex program codes for IRs, which are not specifically defined as dialects in MLIR. 

For WASM-Smith we used two translators and a compiler. The MLIR to LLVM translator rewrites the syntax. While LLVM can be directly generated by MLIR-Forge using the LLVM dialect in MLIR, we opted for translation to make use of our existing prototype. For the LLVM to WASM compilation, we employed \texttt{emcc}, a compiler in the WASM ecosystem. The WASM to WAT translation simply turns bytecode into human-readable text.
Both translators and the compiler are integral components of the system under test, thus of particular interest in testing their correctness and reliability. 

Using this pipeline we perform differential testing using \texttt{emcc} with the \texttt{-O0} and \texttt{-O3} flags. 
Fig.~\ref{fig:5_wasm} showcases one of the bug causing inputs in MLIR for readability.
\texttt{emcc} compiles this input code successfully using \texttt{-O0}, but crashes with \texttt{-O3}. We suspect that a constant propagation pass causes an error when encountering the \texttt{math.fpowi} as it raises zero to the power of zero. Interestingly, if we remove the \texttt{math.absf}, \texttt{emcc} successfully compiles with \texttt{-O3}.
We generated $12,014$ programs over a period of $10$ hours and identified $15$ bug groups.

\subsection{Resources}
\begin{figure}[h]
    \centering
    \begin{subfigure}[b]{0.48\linewidth}
        \centering
        \includegraphics[width=\linewidth]{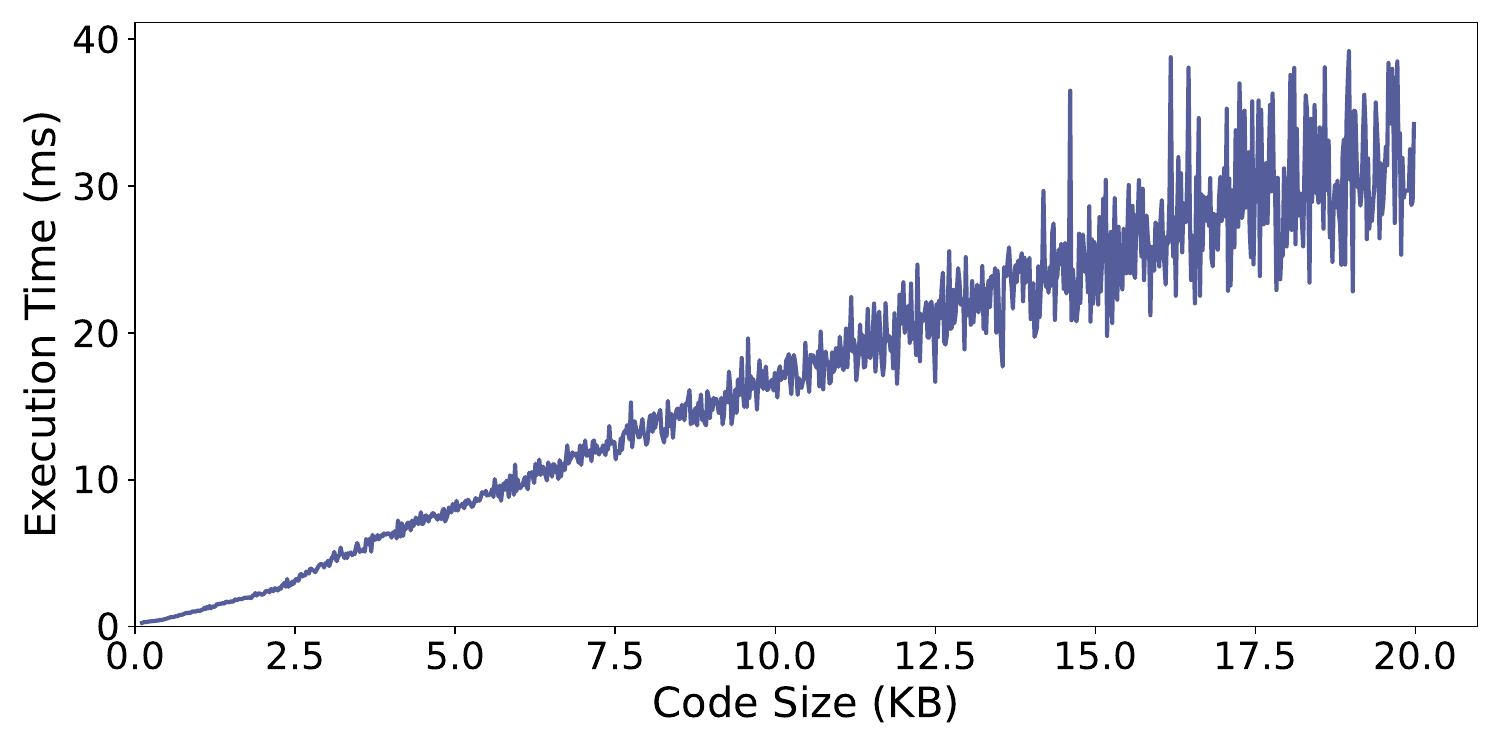}
        \caption{Comparison of code size to execution time.}
        \label{fig:5_time_loc}
    \end{subfigure} 
    \hfill
    \begin{subfigure}[b]{0.48\linewidth}
        \centering
        \includegraphics[width=\linewidth]{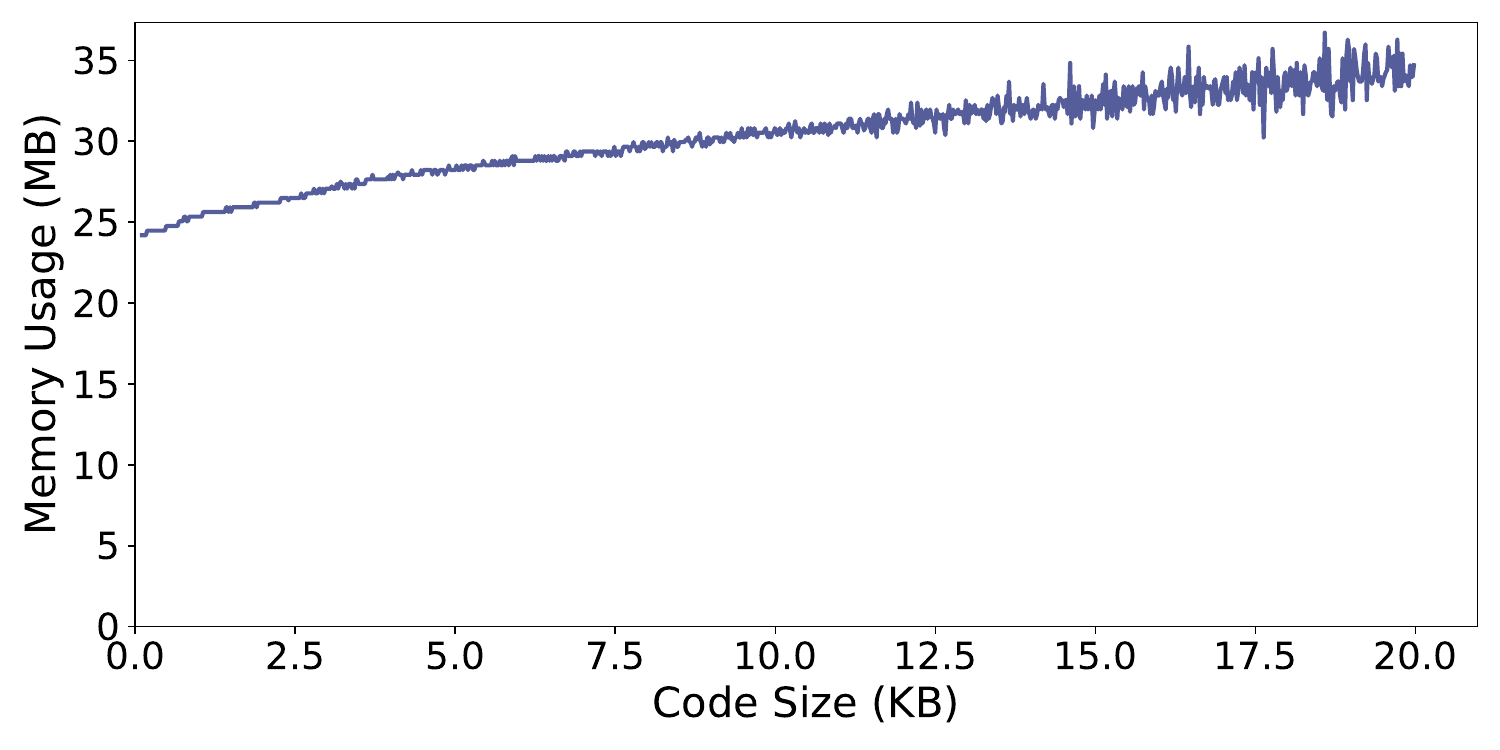}
        \caption{Comparison of code size to memory usage.}
        \label{fig:5_memory_loc}
    \end{subfigure}
    \caption{Measurements of MLIR-Forge's performance.}
    \label{fig:5_time_memory_loc}
\end{figure}

To demonstrate the low time and resources requirements for this type of testing we measure the time spent and the memory consumption during the generation of $100,000$ SDFG-Smith programs, as well as the size of the generated programs.
For our experiment we used a Intel(R) Core(TM) i9-10980XE CPU clocked at $3.00$~GHz.
Fig.~\ref{fig:5_time_memory_loc} depicts the median results grouped into $1,000$ bins.
Note that the generated code size is given in bytes instead of in Lines of Code (LoC), because single line operations may contain an arbitrary number of inputs, increasing the code size in bytes but not in LoC.

As shown in Fig.~\ref{fig:5_time_memory_loc}, MLIR-Forge uses under $36$~MB of memory and completes in less than $40$~ms for programs smaller than $20$~KB, demonstrating how it can easily be used on consumer hardware.
This allows it to be used in rapid testing during fast-iterating development workflows.
Both time and memory consumption appear linearly correlated with the code size.
It is worth noting that MLIR-Forge has not been optimized, frequently performing multiple AST traversals that could be minimized with caching, even further reducing the time and memory consumption for the same workloads.

Highlighting MLIR-Forge's ease of use, each of the random program generators was implemented by a single developer in under a week.
%
To have a tangible measurement of the effort involved in these program generators, we summarize the LoC in the OpGens and TypeGens in Table~\ref{tab:5_loc}. All three random program generators use the arith and math dialects, which have many OpGens with similar structures, such as unary and binary operations. This also demonstrates the code reuse capabilities of MLIR-Forge as the arith and math dialect were written once but used by three different program generators.

\begin{table}[h]
\centering
\caption{Lines of Code (LoC) for OpGens and TypeGens in Smiths}
\label{tab:5_loc}
\begin{tabular}{ccccc}
\textbf{Dialect} & SDFG (Types) & SDFG & Arith & Math \\
\midrule
\textbf{LoC} & 24 & 423 & 1519 & 810 \\
\\
\textbf{Dialect} & Memref (Types) & Memref & SCF & Func \\
\midrule
\textbf{LoC} & 9 & 248 & 179 & 56 \\
\end{tabular}
\end{table}

\section{Related Work}
The landscape of compiler IRs and their associated tools has seen significant advancements in recent years. The development of domain-specific languages, formalization of IR semantics, and the use of machine learning for fuzzing are some of the notable trends in this domain. 

\macsection{Random Program Generators}
In the dynamic realm of software testing, random program generators have emerged as indispensable tools. These generators, by producing a myriad of unpredictable programs, challenge the robustness of compilers, interpreters, and other software systems.
%
Csmith~\cite{10.1145/1993316.1993532} has established itself as a pivotal random program generator, uncovering a multitude of bugs in real-world compilers. Its methodology and success have set a benchmark in the domain of software testing.
%
Cummins et al.~\cite{10.1145/3213846.3213848} took it a step further with DeepSmith, a machine learning tool for compiler testing. Using data from open-source codes, DeepSmith created many realistic programs for testing. After testing for $1,000$~hours, they found issues in every compiler they checked, both commercial and open-source.
%
Similarly, COMFORT~\cite{10.1145/3453483.3454054} brings a fresh perspective to JavaScript (JS) fuzzing by utilizing GPT-2, a deep learning-based language model, to generate valid JS programs. Its primary objective is to identify standard conformance bugs, particularly those aligned with the ECMA-262 (ECMAScript) standard.
EvoSuite~\cite{10.1145/2025113.2025179} and Randoop~\cite{10.1145/1297846.1297902} are notable for their contributions to Java program generation and testing. QuickCheck~\cite{10.1145/357766.351266}, initially developed for Haskell, has been influential in generating test cases for functions across various languages. Tools such as DOMATO~\cite{domato} have been instrumental in generating random JavaScript and web content, respectively, for testing purposes. Skyfire~\cite{7958599} focuses on generating inputs for web browsers, while LangFuzz~\cite{180229} is dedicated to testing the parsing of programming languages.
%
%
%
PolyGlot~\cite{9519403} and other grammar-based generators produce programs independent of the target programming language.
They are well-suited for fuzz testing compiler validators, which prevent the compiler from accepting invalid input. Producing semantically correct code, however, is difficult, which limits their effectiveness in differentially testing compiler optimizations.
%
Xsmith~\cite{10.1145/3624007.3624056} presents a framework to create program generators, prioritizing semantic correctness, which is well-suited for differential testing optimizations, but ineffective for fuzz testing validators. 
%
%
%
%
In contrast, a modular program generator framework, such as MLIR-Forge, bridges a gap between the two techniques by accommodating both testing modes, offering a comprehensive approach to testing novel IRs. Additionally, its capability to generate circular operations, where two operations rely on each other's output, is particularly advantageous for graph-based IRs prevalent in machine learning-centric environments. This capability distinguishes it from other AST construction algorithms, which do not possess this functionality.
MLIR-Forge, offers the same benefits as these tools, but consumes significantly less resources to implement.
This makes it a feasible strategy for testing experimental IRs.

\macsection{Input Fuzzers}
Following the same trend of utilizing randomness, input fuzzing has emerged as a powerful technique for identifying vulnerabilities in software systems. Over the years, a plethora of fuzzing tools have been developed, each with its unique strengths and areas of focus.
%
AFL~\cite{afl} and AFL++~\cite{AFLplusplus-Woot20} are among the most popular coverage-guided fuzzers, using genetic algorithms to efficiently discover new inputs that trigger novel internal states in the target program. Their approach has been foundational in the fuzzing community and has inspired many subsequent tools.
%
LibFuzzer~\cite{libfuzzer} is another coverage-guided fuzzer that operates in-process, repeatedly generating and executing test inputs. Its in-process nature allows for rapid testing cycles, making it particularly effective for certain applications.
%
Honggfuzz~\cite{honggfuzz} is a security-oriented, evolutionary-based fuzzer, emphasizing the discovery of security vulnerabilities. Its approach contrasts with that of Spirv-fuzz~\cite{10.1145/3453483.3454092}, which focuses on generating semantically valid but random shader programs.
%
Other notable fuzzers include Peach Fuzzer~\cite{peachfuzzer}, which can fuzz applications at the protocol level, and Boofuzz~\cite{boofuzz}, a successor of the Sulley~\cite{Sulleyfuzz} fuzzing framework used for network protocol fuzzing. 
%
Radamsa~\cite{radamsa} is a general-purpose fuzzer that can test any program that reads from standard input, while ClusterFuzz~\cite{clusterfuzz} and OSS-Fuzz~\cite{203944} represent scalable fuzzing infrastructures used for large-scale fuzzing campaigns.
%
MLIR-Forge complements fuzzers by making it easy to create test programs for fuzz testing. By providing a rich set of test programs
While tools such as \texttt{spirv-fuzz} change existing tests to find compiler problems, MLIR-Forge can provide a rich set of initial test programs, which enhances the depth and breadth of the testing process.

\macsection{IR Design}
Historically, the design of IRs has been a manual process, making innovation in this area both costly and time-consuming. Developers have traditionally used general-purpose programming languages to design IRs, leading to verbose implementations and making modifications expensive. Significant work has gone into making this process more efficient.
%
As a part of the LLVM~\cite{10.5555/977395.977673} project, MLIR~\cite{mlir} aims to reduce the cost of building domain specific compilers by providing an extensible and reusable infrastructure. MLIR also includes a variety of common tools used in IR developed, such as diagnostics, pass infrastructure, and testing tools.
%
Fehr et al.~\cite{10.1145/3519939.3523700} further simplified the IR development by introducing an IR Definition Language (IRDL), a domain-specific language specifically designed to define IRs.
This language facilitates the implementation of SSA-based IRs.
By enabling a concise and explicit specification of IRs, IRDL lays the groundwork for developing effective tooling to automate the compiler construction process.
%
%
MLIR-Forge complements these efforts by reducing the overhead involved in implementing IR specific testing tools.
There is potential for our testing framework to be integrated directly into the IRDL ecosystem. Such integration would further reduce the manual effort required in the compiler construction process, ensuring that IRs are not only efficiently designed but also rigorously tested.

The works highlighted in this section underscore the diverse advancements in compiler IRs and their associated tools. Our contribution, MLIR-Forge, draws from these innovations, offering a unique perspective in the realm of compiler research. We emphasize the potential of MLIR-Forge to bridge various tools and methodologies, fostering a cohesive compiler research ecosystem.

\section{Conclusion}
With the importance of IR testing to bolster IR research and experimentation, random program generators play an important role.
In this paper we propose a random program generator framework that substantially reduces the time-consuming nature of developing random program generator tools for intermediate representations that serve as testing tools.
This framework divides the process of constructing a random program generator into two distinct tasks: program construction and language specification.

This division reduces the developer's responsibility to providing a concise language specification by defining individual puzzle pieces in the form of individual operations and type generators.
These puzzle pieces specify to the program generator framework how individual program elements can be chained together and how they interact, allowing it to construct random programs without additional information.
Operation and type generators can be shared across languages and representations that share language features, enhancing reuse across different program generators.

Through the development of MLIR-Forge as a proof-of-concept implementation, we demonstrate the practicality of such a framework. We apply it in differential testing with the DaCe parallel programming framework and show its adaptability beyond just MLIR-based IRs by using it to generate WebAssembly programs. All of this is possible with minimal development effort and resource consumption, allowing sophisticated compiler and language testing on consumer hardware.
%
%
MLIR-Forge allows researchers to spend less time on testing and focus more on experimentation. Additionally, we hope to increase trust in new research outcomes in the IR field.


\bibliographystyle{ACM-Reference-Format}
\bibliography{refs}

@article{DBLP:journals/corr/abs-1807-04188,
  author       = {Thierry Moreau and
                  Tianqi Chen and
                  Ziheng Jiang and
                  Luis Ceze and
                  Carlos Guestrin and
                  Arvind Krishnamurthy},
  title        = {{VTA:} An Open Hardware-Software Stack for Deep Learning},
  journal      = {CoRR},
  volume       = {abs/1807.04188},
  year         = {2018},
  url          = {http://arxiv.org/abs/1807.04188},
  eprinttype    = {arXiv},
  eprint       = {1807.04188},
  bibsource    = {dblp computer science bibliography, https://dblp.org}
}

@inproceedings{10.1145/2830772.2830791,
author = {Padmanabha, Shruti and Lukefahr, Andrew and Das, Reetuparna and Mahlke, Scott},
title = {DynaMOS: Dynamic Schedule Migration for Heterogeneous Cores},
year = {2015},
isbn = {9781450340342},
publisher = {Association for Computing Machinery},
address = {New York, NY, USA},
url = {https://doi.org/10.1145/2830772.2830791},
doi = {10.1145/2830772.2830791},
booktitle = {Proceedings of the 48th International Symposium on Microarchitecture},
pages = {322–333},
numpages = {12},
location = {Waikiki, Hawaii},
series = {MICRO-48}
}

@inproceedings{DBLP:journals/corr/abs-1302-6333,
  author       = {Sung{-}Shik T. Q. Jongmans and
                  Farhad Arbab},
  editor       = {Simon J. Gay and
                  Paul Kelly},
  title        = {Modularizing and Specifying Protocols among Threads},
  booktitle    = {Proceedings Fifth Workshop on Programming Language Approaches to Concurrency-
                  and Communication-cEntric Software, {PLACES} 2012, Tallinn, Estonia,
                  31 March 2012},
  series       = {{EPTCS}},
  volume       = {109},
  pages        = {34--45},
  year         = {2012},
  url          = {https://doi.org/10.4204/EPTCS.109.6},
  doi          = {10.4204/EPTCS.109.6},
  bibsource    = {dblp computer science bibliography, https://dblp.org}
}

@InProceedings{10.1007/3-540-46423-9_2,
author="Vall{\'e}e-Rai, Raja
and Gagnon, Etienne
and Hendren, Laurie
and Lam, Patrick
and Pominville, Patrice
and Sundaresan, Vijay",
editor="Watt, David A.",
title="Optimizing Java Bytecode Using the Soot Framework: Is It Feasible?",
booktitle="Compiler Construction",
year="2000",
publisher="Springer Berlin Heidelberg",
address="Berlin, Heidelberg",
pages="18--34",
isbn="978-3-540-46423-5"
}

@inproceedings{mlir,
  author={Lattner, Chris and Amini, Mehdi and Bondhugula, Uday and Cohen, Albert and Davis, Andy and Pienaar, Jacques and Riddle, River and Shpeisman, Tatiana and Vasilache, Nicolas and Zinenko, Oleksandr},
  booktitle={2021 {{IEEE/ACM}} International Symposium on Code Generation and Optimization (CGO)},
  title={{{MLIR}}: Scaling Compiler Infrastructure for Domain Specific Computation},
  year={2021},
  volume={},
  number={},
  pages={2-14},
  doi={10.1109/CGO51591.2021.9370308}
}

@inproceedings{mlir-dace,
    author = {Ben-Nun, Tal and Ates, Berke and Calotoiu, Alexandru and Hoefler, Torsten},
    title = {Bridging Control-Centric and Data-Centric Optimization},
    year = {2023},
    isbn = {9798400701016},
    publisher = {Association for Computing Machinery},
    address = {New York, NY, USA},
    url = {https://doi.org/10.1145/3579990.3580018},
    doi = {10.1145/3579990.3580018},
    booktitle = {Proceedings of the 21st ACM/IEEE International Symposium on Code Generation and Optimization},
    pages = {173–185},
    numpages = {13},
    location = {Montr\'{e}al, QC, Canada},
    series = {CGO 2023}
}

@article{10.1145/3469030,
author = {Gysi, Tobias and M\"{u}ller, Christoph and Zinenko, Oleksandr and Herhut, Stephan and Davis, Eddie and Wicky, Tobias and Fuhrer, Oliver and Hoefler, Torsten and Grosser, Tobias},
title = {Domain-Specific Multi-Level IR Rewriting for GPU: The Open Earth Compiler for GPU-Accelerated Climate Simulation},
year = {2021},
issue_date = {December 2021},
publisher = {Association for Computing Machinery},
address = {New York, NY, USA},
volume = {18},
number = {4},
issn = {1544-3566},
url = {https://doi.org/10.1145/3469030},
doi = {10.1145/3469030},
journal = {ACM Trans. Archit. Code Optim.},
month = {sep},
articleno = {51},
numpages = {23}
}

@misc{chelini2022mom,
      title={MOM: Matrix Operations in MLIR}, 
      author={Lorenzo Chelini and Henrik Barthels and Paolo Bientinesi and Marcin Copik and Tobias Grosser and Daniele G. Spampinato},
      year={2022},
      eprint={2208.10391},
      archivePrefix={arXiv},
      primaryClass={cs.PL}
}

@article{DBLP:journals/corr/abs-2008-08272,
  author       = {Tung D. Le and
                  Gheorghe{-}Teodor Bercea and
                  Tong Chen and
                  Alexandre E. Eichenberger and
                  Haruki Imai and
                  Tian Jin and
                  Kiyokuni Kawachiya and
                  Yasushi Negishi and
                  Kevin O'Brien},
  title        = {Compiling {ONNX} Neural Network Models Using {MLIR}},
  journal      = {CoRR},
  volume       = {abs/2008.08272},
  year         = {2020},
  url          = {https://arxiv.org/abs/2008.08272},
  eprinttype    = {arXiv},
  eprint       = {2008.08272},
  bibsource    = {dblp computer science bibliography, https://dblp.org}
}

@misc{hu2023tpumlir,
      title={TPU-MLIR: A Compiler For TPU Using MLIR}, 
      author={Pengchao Hu and Man Lu and Lei Wang and Guoyue Jiang},
      year={2023},
      eprint={2210.15016},
      archivePrefix={arXiv},
      primaryClass={cs.PL}
}

@misc{mccaskey2021mlir,
      title={A MLIR Dialect for Quantum Assembly Languages}, 
      author={Alexander McCaskey and Thien Nguyen},
      year={2021},
      eprint={2101.11365},
      archivePrefix={arXiv},
      primaryClass={quant-ph}
}

@article{rajamanickam2019software,
  title={Software testing: The generation tools},
  author={Rajamanickam, Leelavathi and Saat, NABM and Daud, SNB},
  journal={Int. J. Adv. Trends Comput. Sci. Eng},
  volume={8},
  number={2},
  pages={231--234},
  year={2019}
}

@inproceedings{10.1145/1993498.1993532,
author = {Yang, Xuejun and Chen, Yang and Eide, Eric and Regehr, John},
title = {Finding and Understanding Bugs in C Compilers},
year = {2011},
isbn = {9781450306638},
publisher = {Association for Computing Machinery},
address = {New York, NY, USA},
url = {https://doi.org/10.1145/1993498.1993532},
doi = {10.1145/1993498.1993532},
booktitle = {Proceedings of the 32nd ACM SIGPLAN Conference on Programming Language Design and Implementation},
pages = {283–294},
numpages = {12},
location = {San Jose, California, USA},
series = {PLDI '11}
}

@article{10.1145/1993316.1993532,
author = {Yang, Xuejun and Chen, Yang and Eide, Eric and Regehr, John},
title = {Finding and Understanding Bugs in C Compilers},
year = {2011},
issue_date = {June 2011},
publisher = {Association for Computing Machinery},
address = {New York, NY, USA},
volume = {46},
number = {6},
issn = {0362-1340},
url = {https://doi.org/10.1145/1993316.1993532},
doi = {10.1145/1993316.1993532},
journal = {SIGPLAN Not.},
month = {jun},
pages = {283–294},
numpages = {12}
}

@inproceedings{10.1145/3213846.3213848,
author = {Cummins, Chris and Petoumenos, Pavlos and Murray, Alastair and Leather, Hugh},
title = {Compiler Fuzzing through Deep Learning},
year = {2018},
isbn = {9781450356992},
publisher = {Association for Computing Machinery},
address = {New York, NY, USA},
url = {https://doi.org/10.1145/3213846.3213848},
doi = {10.1145/3213846.3213848},
booktitle = {Proceedings of the 27th ACM SIGSOFT International Symposium on Software Testing and Analysis},
pages = {95–105},
numpages = {11},
location = {Amsterdam, Netherlands},
series = {ISSTA 2018}
}

@inproceedings{10.1145/3062341.3062363,
author = {Haas, Andreas and Rossberg, Andreas and Schuff, Derek L. and Titzer, Ben L. and Holman, Michael and Gohman, Dan and Wagner, Luke and Zakai, Alon and Bastien, JF},
title = {Bringing the Web up to Speed with WebAssembly},
year = {2017},
isbn = {9781450349888},
publisher = {Association for Computing Machinery},
address = {New York, NY, USA},
url = {https://doi.org/10.1145/3062341.3062363},
doi = {10.1145/3062341.3062363},
booktitle = {Proceedings of the 38th ACM SIGPLAN Conference on Programming Language Design and Implementation},
pages = {185–200},
numpages = {16},
location = {Barcelona, Spain},
series = {PLDI 2017}
}

@article{10.1145/3140587.3062363,
author = {Haas, Andreas and Rossberg, Andreas and Schuff, Derek L. and Titzer, Ben L. and Holman, Michael and Gohman, Dan and Wagner, Luke and Zakai, Alon and Bastien, JF},
title = {Bringing the Web up to Speed with WebAssembly},
year = {2017},
issue_date = {June 2017},
publisher = {Association for Computing Machinery},
address = {New York, NY, USA},
volume = {52},
number = {6},
issn = {0362-1340},
url = {https://doi.org/10.1145/3140587.3062363},
doi = {10.1145/3140587.3062363},
journal = {SIGPLAN Not.},
month = {jun},
pages = {185–200},
numpages = {16}
}

@inproceedings{polygeistPACT,
  title = {Polygeist: Raising C to Polyhedral MLIR},
  author = {Moses, William S. and Chelini, Lorenzo and Zhao, Ruizhe and Zinenko, Oleksandr},
  booktitle = {Proceedings of the ACM International Conference on Parallel Architectures and Compilation Techniques},
  numpages = {12},
  location = {Virtual Event},
  series = {PACT '21},
  publisher = {Association for Computing Machinery},
  year = {2021},
  address = {New York, NY, USA},
  doi={10.1109/PACT52795.2021.00011}
}

@inproceedings{schaad2023fuzzyflow,
author = {Schaad, Philipp and Schneider, Timo and Ben-Nun, Tal and Calotoiu, Alexandru and Ziogas, Alexandros Nikolaos and Hoefler, Torsten},
title = {FuzzyFlow: Leveraging Dataflow To Find and Squash Program Optimization Bugs},
year = {2023},
isbn = {9798400701092},
publisher = {Association for Computing Machinery},
address = {New York, NY, USA},
url = {https://doi.org/10.1145/3581784.3613214},
doi = {10.1145/3581784.3613214},
booktitle = {Proceedings of the International Conference for High Performance Computing, Networking, Storage and Analysis},
articleno = {88},
numpages = {15},
location = {Denver, CO, USA},
series = {SC '23}
}

@inproceedings{10.1145/3519939.3523700,
author = {Fehr, Mathieu and Niu, Jeff and Riddle, River and Amini, Mehdi and Su, Zhendong and Grosser, Tobias},
title = {IRDL: An IR Definition Language for SSA Compilers},
year = {2022},
isbn = {9781450392655},
publisher = {Association for Computing Machinery},
address = {New York, NY, USA},
url = {https://doi.org/10.1145/3519939.3523700},
doi = {10.1145/3519939.3523700},
booktitle = {Proceedings of the 43rd ACM SIGPLAN International Conference on Programming Language Design and Implementation},
pages = {199–212},
numpages = {14},
location = {San Diego, CA, USA},
series = {PLDI 2022}
}

@inproceedings{10.1145/3453483.3454054,
author = {Ye, Guixin and Tang, Zhanyong and Tan, Shin Hwei and Huang, Songfang and Fang, Dingyi and Sun, Xiaoyang and Bian, Lizhong and Wang, Haibo and Wang, Zheng},
title = {Automated Conformance Testing for JavaScript Engines via Deep Compiler Fuzzing},
year = {2021},
isbn = {9781450383912},
publisher = {Association for Computing Machinery},
address = {New York, NY, USA},
url = {https://doi.org/10.1145/3453483.3454054},
doi = {10.1145/3453483.3454054},
booktitle = {Proceedings of the 42nd ACM SIGPLAN International Conference on Programming Language Design and Implementation},
pages = {435–450},
numpages = {16},
location = {Virtual, Canada},
series = {PLDI 2021}
}

@inproceedings{10.1145/3453483.3454092,
author = {Donaldson, Alastair F. and Thomson, Paul and Teliman, Vasyl and Milizia, Stefano and Maselco, Andr\'{e} Perez and Karpi\'{n}ski, Antoni},
title = {Test-Case Reduction and Deduplication Almost for Free with Transformation-Based Compiler Testing},
year = {2021},
isbn = {9781450383912},
publisher = {Association for Computing Machinery},
address = {New York, NY, USA},
url = {https://doi.org/10.1145/3453483.3454092},
doi = {10.1145/3453483.3454092},
booktitle = {Proceedings of the 42nd ACM SIGPLAN International Conference on Programming Language Design and Implementation},
pages = {1017–1032},
numpages = {16},
location = {Virtual, Canada},
series = {PLDI 2021}
}

@inproceedings{Muller:HASE05,
  title = {A Framework for Simplifying the Development of Kernel
	  Schedulers: Design and Performance Evaluation},
  author = {Gilles Muller and Julia L. Lawall and Herv\'e Duchesne},
  booktitle = {HASE 2005 - High Assurance Systems Engineering Conference},
  month = oct,
  year = 2005,
  address = {Heidelberg, Germany},
  pages = {56--65}
}

@inproceedings {AFLplusplus-Woot20,
author = {Andrea Fioraldi and Dominik Maier and Heiko Ei{\ss}feldt and Marc Heuse},
title = {{AFL++}: Combining Incremental Steps of Fuzzing Research},
booktitle = {14th {USENIX} Workshop on Offensive Technologies ({WOOT} 20)},
year = {2020},
publisher = {{USENIX} Association},
month = aug,
}

@misc{afl,
    title={AFL: American Fuzzy Lop},
    year={2023},
    url={https://github.com/google/AFL},
    author={Michal Zalewski}
}

@misc{libfuzzer,
    title={libFuzzer: In-process, coverage-guided fuzzing},
    year={2023},
    url={https://llvm.org/docs/LibFuzzer.html},
    author={LLVM Project}
}

@misc{honggfuzz,
    title={Honggfuzz: Evolutionary-based fuzzer},
    year={2023},
    url={https://github.com/google/honggfuzz},
    author={Google}
}

@misc{peachfuzzer,
    title={Peach Fuzzer: Protocol-level fuzzing},
    year={2023},
    url={https://peachtech.gitlab.io/peach-fuzzer-community},
    author={Michael Eddington}
}

@misc{boofuzz,
    title={Boofuzz: Network protocol fuzzing},
    year={2023},
    url={https://github.com/jtpereyda/boofuzz},
    author={Joshua Pereyda}
}

@misc{Sulleyfuzz,
    title={Sulley: An unattended fuzzing framework},
    year={2023},
    url={https://github.com/OpenRCE/sulley},
    author={OpenRCE}
}

@misc{radamsa,
    title={Radamsa: General-purpose fuzzer},
    year={2023},
    url={https://gitlab.com/akihe/radamsa},
    author={Aki Helin}
}

@misc{clusterfuzz,
    title={ClusterFuzz: Scalable fuzzing infrastructure},
    year={2023},
    url={https://github.com/google/clusterfuzz},
    author={Google}
}

@misc{domato,
    title={DOMATO: DOM Fuzzer},
    year={2023},
    url={https://github.com/googleprojectzero/domato},
    author={Ivan Fratric}
}

@conference {203944,
author = {Kostya Serebryany},
title = {{OSS-Fuzz} - Google{\textquoteright}s continuous fuzzing service for open source software},
year = {2017},
address = {Vancouver, BC},
publisher = {USENIX Association},
month = aug
}

@inproceedings{10.1145/2025113.2025179,
author = {Fraser, Gordon and Arcuri, Andrea},
title = {EvoSuite: Automatic Test Suite Generation for Object-Oriented Software},
year = {2011},
isbn = {9781450304436},
publisher = {Association for Computing Machinery},
address = {New York, NY, USA},
url = {https://doi.org/10.1145/2025113.2025179},
doi = {10.1145/2025113.2025179},
booktitle = {Proceedings of the 19th ACM SIGSOFT Symposium and the 13th European Conference on Foundations of Software Engineering},
pages = {416–419},
numpages = {4},
location = {Szeged, Hungary},
series = {ESEC/FSE '11}
}

@inproceedings{10.1145/1297846.1297902,
author = {Pacheco, Carlos and Ernst, Michael D.},
title = {Randoop: Feedback-Directed Random Testing for Java},
year = {2007},
isbn = {9781595938657},
publisher = {Association for Computing Machinery},
address = {New York, NY, USA},
url = {https://doi.org/10.1145/1297846.1297902},
doi = {10.1145/1297846.1297902},
booktitle = {Companion to the 22nd ACM SIGPLAN Conference on Object-Oriented Programming Systems and Applications Companion},
pages = {815–816},
numpages = {2},
location = {Montreal, Quebec, Canada},
series = {OOPSLA '07}
}

@article{10.1145/357766.351266,
author = {Claessen, Koen and Hughes, John},
title = {QuickCheck: A Lightweight Tool for Random Testing of Haskell Programs},
year = {2000},
issue_date = {Sept. 2000},
publisher = {Association for Computing Machinery},
address = {New York, NY, USA},
volume = {35},
number = {9},
issn = {0362-1340},
url = {https://doi.org/10.1145/357766.351266},
doi = {10.1145/357766.351266},
journal = {SIGPLAN Not.},
month = {sep},
pages = {268–279},
numpages = {12}
}

@INPROCEEDINGS{7958599,
  author={Wang, Junjie and Chen, Bihuan and Wei, Lei and Liu, Yang},
  booktitle={2017 IEEE Symposium on Security and Privacy (SP)}, 
  title={Skyfire: Data-Driven Seed Generation for Fuzzing}, 
  year={2017},
  volume={},
  number={},
  pages={579-594},
  doi={10.1109/SP.2017.23},
  ISSN={2375-1207},
  month={May},}

@inproceedings {180229,
author = {Christian Holler and Kim Herzig and Andreas Zeller},
title = {Fuzzing with Code Fragments},
booktitle = {21st USENIX Security Symposium (USENIX Security 12)},
year = {2012},
isbn = {978-931971-95-9},
address = {Bellevue, WA},
pages = {445--458},
url = {https://www.usenix.org/conference/usenixsecurity12/technical-sessions/presentation/holler},
publisher = {USENIX Association},
month = aug
}

@inproceedings{10.5555/977395.977673,
author = {Lattner, Chris and Adve, Vikram},
title = {LLVM: A Compilation Framework for Lifelong Program Analysis \& Transformation},
year = {2004},
isbn = {0769521029},
publisher = {IEEE Computer Society},
address = {USA},
booktitle = {Proceedings of the International Symposium on Code Generation and Optimization: Feedback-Directed and Runtime Optimization},
pages = {75},
location = {Palo Alto, California},
series = {CGO '04}
}

@inproceedings{dace,
  author    = {Ben-Nun, Tal and de~Fine~Licht, Johannes and Ziogas, Alexandros Nikolaos and Schneider, Timo and Hoefler, Torsten},
  title     = {Stateful Dataflow Multigraphs: A Data-Centric Model for Performance Portability on Heterogeneous Architectures},
  year      = {2019},
  booktitle = {Proceedings of the International Conference for High Performance Computing, Networking, Storage and Analysis},
  series = {SC '19}
}

@inproceedings{10.1145/2048147.2048224,
author = {Zakai, Alon},
title = {Emscripten: An LLVM-to-JavaScript Compiler},
year = {2011},
isbn = {9781450309424},
publisher = {Association for Computing Machinery},
address = {New York, NY, USA},
url = {https://doi.org/10.1145/2048147.2048224},
doi = {10.1145/2048147.2048224},
booktitle = {Proceedings of the ACM International Conference Companion on Object Oriented Programming Systems Languages and Applications Companion},
pages = {301–312},
numpages = {12},
location = {Portland, Oregon, USA},
series = {OOPSLA '11}
}

@inproceedings{10.1145/3624007.3624056,
author = {Hatch, William and Darragh, Pierce and Porncharoenwase, Sorawee and Watson, Guy and Eide, Eric},
title = {Generating Conforming Programs with Xsmith},
year = {2023},
isbn = {9798400704062},
publisher = {Association for Computing Machinery},
address = {New York, NY, USA},
url = {https://doi.org/10.1145/3624007.3624056},
doi = {10.1145/3624007.3624056},
booktitle = {Proceedings of the 22nd ACM SIGPLAN International Conference on Generative Programming: Concepts and Experiences},
pages = {86–99},
numpages = {14},
location = {Cascais, Portugal},
series = {GPCE 2023}
}

@INPROCEEDINGS{9519403,
  author={Chen, Yongheng and Zhong, Rui and Hu, Hong and Zhang, Hangfan and Yang, Yupeng and Wu, Dinghao and Lee, Wenke},
  booktitle={2021 IEEE Symposium on Security and Privacy (SP)}, 
  title={One Engine to Fuzz ’em All: Generic Language Processor Testing with Semantic Validation}, 
  year={2021},
  volume={},
  number={},
  pages={642-658},
  doi={10.1109/SP40001.2021.00071}}

\appendix

\end{document}